\documentstyle[12pt]{article}
\def\rbf{{\bf r}}
\def\pbf{{\bf p}}
\def\prbf{{\bf p},{\bf r}}
\begin{document}
\input epsf
\begin{center}
{\bf
A Study of Bose-Einstein Condensation in a Two-Dimensional 
Trapped Gas
\vspace*{0.25cm}\\}
{W. J. Mullin\\}
{\it Department of Physics and Astronomy, University of
Massachusetts,\\
Amherst, MA 01003-3720, USA\vspace*{0.25cm}\\}

\vskip 24pt
\end{center}

\centerline{\bf Abstract}
\vskip 12pt \baselineskip 16pt
We examine the possibility of Bose-Einstein condensation (BEC) in 
two-dimensional (2D) system of interacting particles in a trap.  We 
use a self-consistent mean-field theory of Bose particles interacting 
by a contact interaction in the Popov and WKB approximations.  The 
equations show that the normal state has a phase transition at some 
critical temperature \( T_{c} \) but below \( T_{c} \) the 
Bose-Einstein condensed state is {\em not} a consistent solution of 
the equations in the thermodynamic limit.  This result agrees with a 
theorem recently discussed by the author that shows that a BEC state 
is impossible for an interacting gas in a 2D trap in the thermodynamic 
limit.

PACS numbers: 03.75.Fi, 05.30.Jp, 05.70.Fh, 32.80.Pj, 67.40.Db
\vskip 12pt

\baselineskip 20pt

\centerline{\bf I.~~Introduction}
\vskip 12pt
Recent experiments on alkali atoms\cite{exps} confined in 
three-dimensional (3D) magnetic traps and cooled by evaporation 
techniques have led to the observation of Bose-Einstein condensation 
(BEC).  In these experiments the number of particles cooled has ranged 
from several thousand to millions.  Experiments to date have naturally 
been in 3D systems.  If experimentally it becomes possible to make one 
dimension of the trap so narrow that the harmonic states are very 
greatly separated, then the system could be a reasonable simulation of 
a two-dimensional system.  Further, there is the possibility of an 
adsorbed gas, such as spin-polarized hydrogen on liquid 
helium,\cite{Sil} forming a 2D system.

The ideal Bose gas in two dimensions trapped in a harmonic potential 
{\em has} a Bose-Einstein condensation (BEC).\cite{Mull,Mull2} 
However, the author\cite{Mull2} recently demonstrated how theorems by 
Hohenberg and Chester could be used to show that the trapped system 
can have no BEC, in the thermodynamic limit, if there are interactions 
that prevent the density from diverging at any position.  The ideal 
gas has a divergence at the origin at and below the transition 
temperature, which excludes it from application of the theorem, and a 
BEC does occur. 

Our intent here is to test the general theorem by an explicit 
mean-field computation of the gas properties when interactions are 
present.  We consider particles interacting by a contact potential.  
In 3D such an interaction is a pseudo-potential for a hard-core 
interaction.  This is not the case in 2D, however it provides a 
simple model that can be analyzed by mean-field theory.

Recently Griffin\cite{Grif} discussed the generalized 
finite-temperature Hartree-Fock-Bogoliubov (HFB) equations of motion 
for a Bose gas in a nonuniform potential and the various 
approximations for treating it.  He points out that a particularly 
useful gapless approximation is that of Popov.  Giorgini {\em et al} 
(GPS)\cite{Gps} have solved the equations in this form for three 
dimensions (3D) by use of the WKB approximation.  Hutchinson {\em et 
al} \cite{Grif2} have solved the equations directly.  

We study the 2D system in the thermodynamic limit where the condensate 
equation may be treated exactly by the Fermi-Thomas approach.  As in 
the ideal gas case, in the equations without condensate, there is a 
critical temperature below which no value of the chemical potential 
satisfies the condition on the number of particles.  At that point one 
usually invokes the presence of a condensate to satisfy the 
particle-number condition.  However, we find that {\em there is no 
solution to the self-consistent equations for condensate and 
excited-state particles below the critical temperature}.  Apparently 
the noncondensed system becomes unstable at the critical 
temperature and makes a transition to some other state, but the state 
is {\em not} the BEC state.  Possibly there is a Kosterlitz-Thouless 
transition,\cite{Sil} but we have not yet checked that 
hypothesis.\cite{Shev}

\vskip 12pt
\centerline{\bf II.~~THE THERMODYNAMIC LIMIT FOR HARMONIC TRAPS}
\vskip 12pt

Consider a 2D system of N Bose particles in a spherically symmetric 
harmonic trap.  The harmonic potential is given by
\begin{equation}
U(\rbf)=\frac{1}{2}U_{0}(\frac{r}{R})^{2}  \label{pot}
\end{equation}
where \thinspace $r^{2}=x^{2}+y^{2}$ and $R$ is a range parameter that will
be handy for taking the thermodynamic limit. The angular frequency is 
\begin{equation}
\omega =\sqrt{\frac{U_{0}}{R^{2}m}}  \label{omeg}
\end{equation}
where $m$ is the particle mass. 

If we wish to take the thermodynamic limit we must increase the 
``volume'' while keeping the average density fixed.  The average 
density is proportional to $\rho =N/R^{2}$ where $R$ is the range 
parameter in Eq.~(\ref {pot}) chosen at some convenient temperature to 
be a distance within which the majority of particles resides.  
Increasing the volume then implies weakening the potential, by 
increasing \( R \), while $N$ increases.  From Eq.~(\ref{omeg}) we 
see that this requires keeping $N\omega^{2}= const$ while 
$N\rightarrow \infty $ , This limiting process has been considered 
previously.\cite{Mull,Mull2,Damle} Define a characteristic temperature
\begin{equation}
T_{0}=\frac{\hbar}{k_{B}}\sqrt{\frac{U_{0}\rho }{m}}  \label{T0}
\end{equation}
where $k_{B}$ is the Boltzmann constant.  We see that 
$k_{B}T_{0}=\sqrt{N} \hbar \omega $ remains constant as the 
thermodynamic limit is taken.  For the ideal gas in a harmonic trap 
there is a phase transition\cite {Mull,Mull2} at 
$T_{c}=T_{0}/\sqrt{\zeta (2)}$ 
where $\zeta (\sigma )$ is the Riemann $\zeta $ -function.

\vskip 12pt
\centerline{\bf III.~~HARTREE-FOCK-BOGOLIUBOV EQUATIONS}
\vskip 12pt

Recently Griffin\cite{Grif} has discussed the derivation of a 
self-consistent mean-field treatment of the inhomogeneous interacting 
Bose gas valid at finite temperatures.  One writes the field operator 
for the bosons as
\begin{equation}
\hat{\psi}=\Phi +\tilde{\psi}  \label{wave}
\end{equation}
where $\Phi =\left\langle \hat{\psi}\right\rangle $ is the condensate 
wave function and $\tilde{\psi}$ describes fluctuations.  What results 
in the Popov approximation is a generalized Gross-Pitevskii equation 
valid at $T>0$ that now depends not only on the local condensate 
density $n_{0}(\rbf)=\left\langle \Phi ^{*}\Phi \right\rangle $ 
but also on the density $n_{T}(\rbf)=\left\langle \tilde{ 
\psi}^{\dagger }\tilde{\psi}\right\rangle $ of particles in excited 
states.  When we consider a contact interaction of strength $g$, the 
equation for the condensate is\cite{Grif}
\begin{equation}
\left[ \Lambda -gn_{0}(\rbf)\right] \Phi =0  \label{cond}
\end{equation}
where the operator $\Lambda $ is 
\begin{equation}
\Lambda =-\frac{\hbar ^{2}}{2m}\nabla ^{2}+U(\rbf)-
\mu +2gn(\rbf),  \label{oper}
\end{equation}
$\mu $ is the chemical potential, and $n=n_{0}+n_{T}$ is the total 
local density. 

There is also an equation for $\tilde{\psi}$ that depends on 
$n_{0}$ and $n_{T}$.  A Bogoliubov transformation of this equation 
leads to a pair of differential equations, which Hutchinson {\em et 
al} \cite {Grif2} have solved in 3D by introducing an eigenfunction 
basis.  \ Giorgini {\em et al} \cite{Gps} have used the WKB 
approximation to simplify and solve the equations for the excitations 
in 3D.  The result of the same procedure in 2D is the following: The 
excitation spectrum is
\begin{equation}
\epsilon (\prbf)=\sqrt{\bar{\Lambda}^{2}-(gn_{0})^{2}}  
\label{excit}
\end{equation}
where $\bar{\Lambda}=\frac{p^{2}}{2m}+U(\rbf)- \mu 
+2gn(\rbf)$. The density of the excited particles is given by
\begin{equation}
n_{T}(\rbf)=\frac{1}{h^{2}}\int d\pbf\left\{ \left[ 
u^{2}( \prbf)+v^{2}(\prbf)\right] 
f(\prbf)+v^{2}(\prbf)\right\} \label{nT}
\end{equation}
with 
\begin{equation}
f(\prbf)=\frac{1}{e^{\beta \epsilon }-1}  \label{dist}
\end{equation}
and 
\begin{equation}
u^{2}(\prbf)=\frac{\bar{\Lambda}+\epsilon }{2\epsilon }  \label{u}
\end{equation}
\begin{equation}
v^{2}(\prbf)=\frac{\bar{\Lambda}-\epsilon }{2\epsilon }  \label{v}
\end{equation}
The total number of particles satisfies 
\begin{equation}
\label{Ncond}
N=\int d\rbf\left[ n_{0}(\rbf)+n_{T}(\rbf)\right] 
\end{equation}
The semiclassical approach makes sense if $k_{B}T\gg \hbar \omega$.

\vskip 12pt
\centerline{\bf IV.~~A TRANSITION, BUT TO WHAT STATE?}
\vskip 12pt

In 2D Eq.~(\ref{nT}) can be integrated.  When one changes 
variables from $p\bf{\,}$ to $y=\beta \epsilon $ the integral takes 
on a particularly simple form:
\begin{eqnarray}
n_{T}(\rbf) &=&\frac{1}{\lambda 
^{2}}\int_{\sqrt{t^{2}-s^{2}}}^{\infty }dy\left[ 
\frac{1}{e^{y}-1}+\frac{1}{2}\left( 1-\frac{y}{\sqrt{y^{2}+s^{2}}}%
\right) \right] \nonumber \\ &=&\frac{1}{\lambda ^{2}}\left\{ -\ln 
\left[ 1-\exp (-\sqrt{t^{2}-s^{2}})\right] 
+t-\sqrt{t^{2}-s^{2}}\right\} \label{nTint}
\end{eqnarray}
where $t=\beta \left( U(\rbf)-\mu +2gn(\rbf)\right) $,  
$s=\beta gn_{0}(\rbf)$, and $\lambda^{2}=h^{2}/2\pi m k_{B}T$.

When there is no condensate the density is simply 
\begin{equation}
n_{T}^{(>)}(\rbf)=-\frac{1}{\lambda ^{2}}\ln \left[ 1-\exp (-t)\right] 
\label{nTabv}
\end{equation}
We have 
solved Eq.~(\ref{nTabv}) self-consistently for $\mu $ by numerical 
means and find that, just as in the ideal gas case, there is a 
solution only for $T$ greater that some critical value.  Sample 
results are shown in Fig.~1.  Below the critical temperature we expect 
a new phase to exist.  To see if the transition is to the BEC state we 
must consider the full set of equations.

Baym and Pethick\cite{BP} have shown that the Fermi-Thomas 
approximation in which the kinetic energy is neglected is valid when 
$N$ is large.  In our 2D case one can show that the kinetic 
energy diminishes relatively as $1/N$.  Thus in the 
thermodynamic limit the Fermi-Thomas approximation is exact and 
Eq.~(\ref{cond}) leads to
\begin{equation}
n_{0}=\frac{1}{g}\left[ \mu-U-2gn_{T} \right]   \label{n0FT}
\end{equation}
This equation along with Eqs.~(\ref{Ncond}) and (\ref{nTint}) must be 
solved self-consistently.  However, we see immediately that this is 
impossible because from Eq.~(\ref{n0FT}), $s=t,$ and the exponential 
in Eq.~(\ref{nTint}) vanishes giving nonsense for all conditions.  
What has happened is that the lower limit in the integral form of 
Eq.~(\ref{nTint}) has vanished indicating that the long-wavelength 
phonons have destabilized long-range order just as in the homogeneous 
case.  The BEC equations are inconsistent in agreement with the 
theorem discussed in Ref.~4, and there is no BEC in 2D in the 
thermodynamic limit.  Note however that the mean-field equations for 
high temperature predict that there {\em is} a phase transition at 
some critical temperature.  If those equations have any validity in 
describing a real system they tell us that the normal state of the 
interacting gas becomes unstable at some temperature.  However, what 
state becomes stable is not apparent from the present discussion,  
possibly a Kosterlitz-Thouless transition occurs.\cite{Shev}

Real experiments are not done in the thermodynamic limit but with a 
finite number of particles.  There can be a pseudo-condensation, a 
macroscopic number of particles in the lowest state below some 
temperature that would go to zero in the thermodynamic 
limit.\cite{Mull2,KvD} Experiments in which this ``transition 
temperature'' is tracked as a function of \( N \) might be 
possible.  Further theoretical computations for finite \( N \) 
are in order.

\vskip 12pt
\centerline{\bf Figure Caption}
\vskip 12pt
Fig. 1. \( \alpha = -\mu/k_{B}T \) versus temperature divided by \( 
T_{0} \) for a sample set of parameters.  Solid line: interacting gas 
with parameters $\gamma (\equiv g\rho^{2}/k_{B}T_{0})=1$ and \( 
\tau_{0}(\equiv k_{B}T_{0}\rho^{2}m/\hbar^{2})=1 \).  Dotted line: 
Non-interacting case (\( \gamma=0, \tau_{0}=1 \)).  With interactions 
there is a solution for \( \mu \) above a critical reduced 
temperature, \( \tau_{c} \), but below, the self-consistent equations 
have no solution indicating there is no long-range condensate order.


\begin{thebibliography}{9}

\bibitem{exps}  M. H. Anderson, J. R. Ensher, M. R. Matthews, C. E. Wieman,
and E. A. Cornell, {\it Science}, {\bf 29}, 198 (1995);
K. B. Davis, M.-O. Mewes, N. J. van Druten, D. S. Durfee, D.
M. Kurn, and W. Ketterle, {\it Phys. Rev. Lett.} {\bf 75}, 3969 (1995);
C. C. Bradley, C. A. Sacket, and R. G. Hulet, {\it Phys.
Rev. Lett.} {\bf 78}, 985 (1997).

\bibitem{Sil} I. F. Silvera, in {\it Bose-Einstein Condensation},
ed. A. Griffin, D. W. Snoke, and S. Stringari (Cambridge University Press,
Cambridge, UK, 1995) p. 160.

\bibitem{Mull} R. Masut and W. J. Mullin, {\it Amer. J. Phys.} 
{\bf 47}, 493 (1979).

\bibitem{Mull2}  W. J. Mullin, {\it J. Low Temp. Phys}. {\bf 106}, 615
(1997) and references therein.

\bibitem{Grif}  A. Griffin, {\it Phys. Rev. B} {\bf 53}, 9341 (1996).

\bibitem{Gps}  S. Giorgini, L. P. Pitaevskii, and S. Stringari {\it Phys. 
Rev. A} {\bf 54}, R4633 (1996); {\it Phys. Rev. Lett.} {\bf 78}, 
3987 (1997); preprint.

\bibitem{Grif2} D. A. Hutchinson, E. Zaremba, and A. Griffin, 
cond-mat/9611023.

\bibitem{Shev}S. I. Shevchenkov [Sov. Phys. JETP {\bf 73}1009 
(1991); Sov. J. Low Temp.  Phys. {\bf 18}, 223 
(1992)], using a different analysis, also concluded that a 2D BEC was not 
possible, but that a KT transition can occur.  Discussion of 
these papers is deferred to a future publication.

\bibitem{Damle} K. Damle, T. Senthil, S. N. Majumdar, and S. 
Sachdev, {\it Europhys. Lett.} {\bf 36}, 6 (1996).

\bibitem{BP}  G. Baym and C. Pethick, {\it Phys. Rev. Lett.} 
{\bf 76}, 6 (1996).

\bibitem{KvD}  W. Ketterle and N. J. van Druten, {\it Phys. Rev. A} 
{\bf 54}, 656 (1996).

\end{thebibliography}
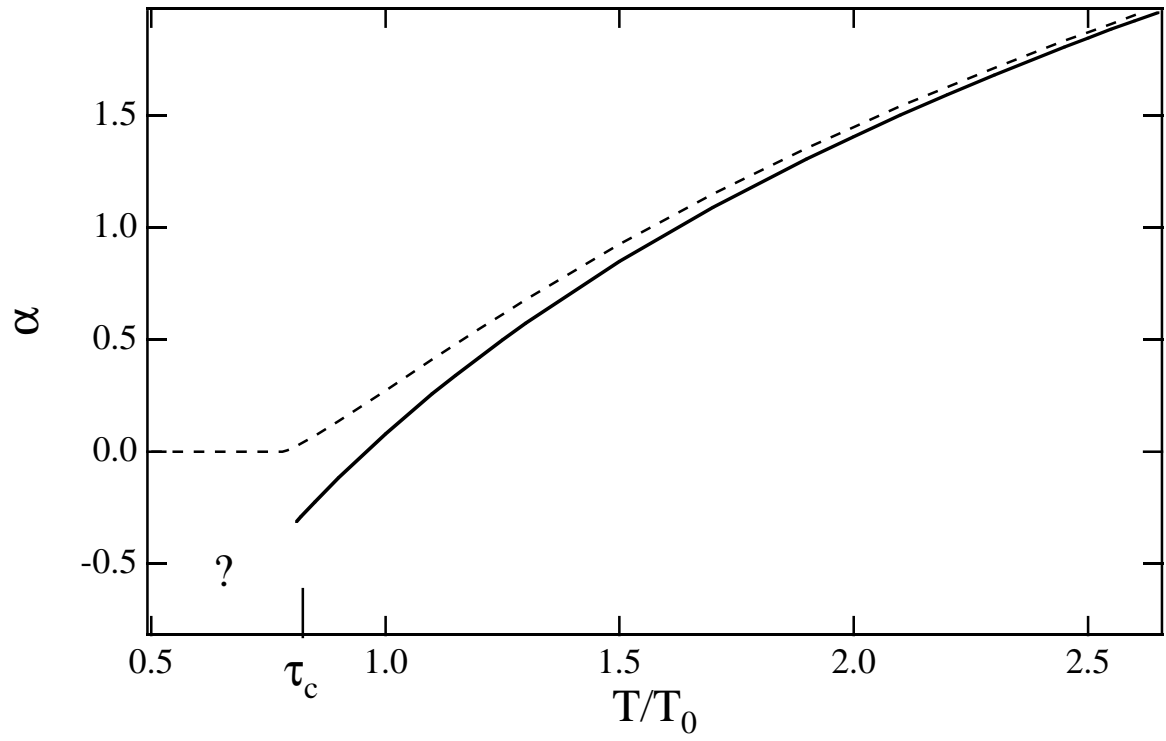
\end{document}